\documentclass[prl,aps,amssymb,superscriptaddress,twocolumn,10pt]{revtex4}
\usepackage{graphicx}
\usepackage{amsmath}

\newcommand{\unit}[1]{\mathrm{~#1}}
\newcommand{\guil}[1]{``#1''}
\newcommand{\moy}[1]{\left < #1 \right >}
\newcommand{\inp}[1]{\left(#1\right)}
\newcommand{\insb}[1]{\left[#1\right]}

\newcommand{\ee}[1]{\times 10^{#1}}
\newcommand{\vdc}{V_{\mathrm{dc}}}
\newcommand{\vac}{V_{\mathrm{ac}}}

\begin{document}

\title{Experimental Violation Of Bell-like Inequalities By Electronic Shot Noise}

\author{Jean-Charles Forgues}\email{jean-charles.forgues@usherbrooke.ca}
\author{Christian Lupien}\email{christian.lupien@usherbrooke.ca}
\author{Bertrand Reulet}\email{bertrand.reulet@usherbrooke.ca}
\affiliation{D\'{e}partement de Physique, Universit\'{e} de Sherbrooke, Sherbrooke, Qu\'{e}bec, Canada, J1K 2R1}

\date{\today}
\begin{abstract}
We report measurements of the correlations between electromagnetic field quadratures at two frequencies $f_1=7\unit{GHz}$ and $f_1=7.5\unit{GHz}$ of the radiation emitted by a tunnel junction placed at very low temperature and excited at frequency $f_1+f_2$. We demonstrate the existence of two-mode squeezing and violation of a Bell-like inequality, thereby proving the existence of entanglement in the quantum shot noise radiated by the tunnel junction.
\end{abstract}

\maketitle
Electrical current flowing in a conductor always fluctuates in time, a phenomenon usually referred to as \guil{electrical noise}. At room temperature, one way this can be described is using a time-dependent current $I(t)$. While the dc current corresponds to the average $\moy{I(t)}$, current fluctuations are characterized by their statistical properties such as their second order correlator $\moy{I(t)I(t')}$, where the brackets $\moy{.}$ represent the statistical average. An alternate approach is to consider that this time-dependent current generates a random electromagnetic field that propagates along the electrical wires. Both these descriptions are equivalent. For example, the equilibrium current fluctuations (Johnson-Nyquist noise \cite{Johnson1928, Nyquist1928}) correspond to the blackbody radiation in one dimension\cite{Oliver1965}. More precisely, the power radiated by a sample at frequency $f$ in a cable is proportional to the spectral density $S(f)$ of current fluctuation which, at high temperature and at equilibrium (i.e. with no bias), is given by $S_{0}(hf\ll k_BT)=2k_BT/R$ where $T$ is the temperature and $R$ the electrical resistance of the sample\cite{Callen1951}.

In short samples at very low temperatures, electrons obey quantum mechanics. Thus, electron transport can no longer be modeled by a time-dependent, classical number $I(t)$, but needs to be described by an operator $\hat I(t)$. Current fluctuations are characterized by correlators such as $\moy{\hat I(t)\hat I(t')}$. Quantum predictions differ from classical ones only when the energy $hf$ associated with the electromagnetic field is comparable with energies associated with the temperature $k_BT$ and the voltage $eV$. Hence for $hf\gg k_BT,eV$, the thermal energy $k_BT$ in the expression of $S_{0}(f)$ has to be replaced by that of vacuum fluctuations, $hf/2$. Some general link between the statistics of current fluctuations and that of the detected electromagnetic field is required beyond the correspondence between spectral density of current fluctuations and radiated power \cite{Beenaker2001,Beenaker2004,Lebedev2010,Qassemi}. In particular, since the statistics of current fluctuations can be tailored by engineering the shape of the time-dependent bias voltage \cite{Gabelli2013}, it may be possible to induce non-classical correlations in the electromagnetic field generated by a quantum conductor. For example, an ac bias at frequency $f_0$ generates correlations between current fluctuations at frequencies $f_1$ and $f_2$, i.e. $\moy {\hat I(f_1)\hat I(f_2)}\neq 0$, if $f_1\pm f_2=nf_0$ with $n$ integer \cite{GR1,GR2,GR3}. This is responsible for the existence of correlated power fluctuations \cite{C4classique} and for the emission of photon pairs \cite{C4quantique} recently observed. For $f_1=f_2$, $\moy{\hat I^2(f_1)}\neq 0$ leads to vacuum squeezing \cite{BBRB,PAN_squeezing}.
In this article, we report measurements of the correlations between electromagnetic field quadratures at frequencies $f_1$ and $f_2$ when the sample is irradiated at frequency $f_0=f_1+f_2$.  By analyzing these correlations, we show that the electromagnetic field produced by electronic shot noise can be described in a way similar to an Einstein-Podolski-Rosen (EPR) pair: when measuring fluctuations at only one frequency, i.e. one mode of the electromagnetic field, no quadrature is preferred. But when measuring two modes, we observe strong correlations between identical quadratures. These correlations are stronger than what is allowed by classical mechanics as proven by their violation of Bell-like inequalities.

Entanglement of photons of different frequencies has already been observed in superconducting devices engineered for that purpose in Refs. \cite{Eichler, Flurin, Nguyen}, where frequencies $f_1$ and $f_2$ are fixed by resonators and the entanglement comes from a non-linear element, a Josephson junction. What we show here is that any quantum conductor excited at frequency $f_0$ emits entangled radiation at \emph{any} pair of frequencies $f_1$, $f_2$ such that $f_0=f_1+f_2$. This property is demonstrated using a tunnel junction but our results clearly stand for any device that exhibits quantum shot noise.
The key ingredient for the appearance of entanglement is the following: noise at any frequency $f_1$ modulated by an ac voltage at frequency $f_0$ gives rise to sidebands with a well defined phase. These sidebands, located at frequencies $\pm f_1\pm nf_0$ with $n$ integer, are correlated with the current fluctuations at frequency $f_0$. The particular case $f_2=-f_1+f_0$ we study here corresponds to the maximum correlation.

\begin{figure}
    \includegraphics[width=\columnwidth] {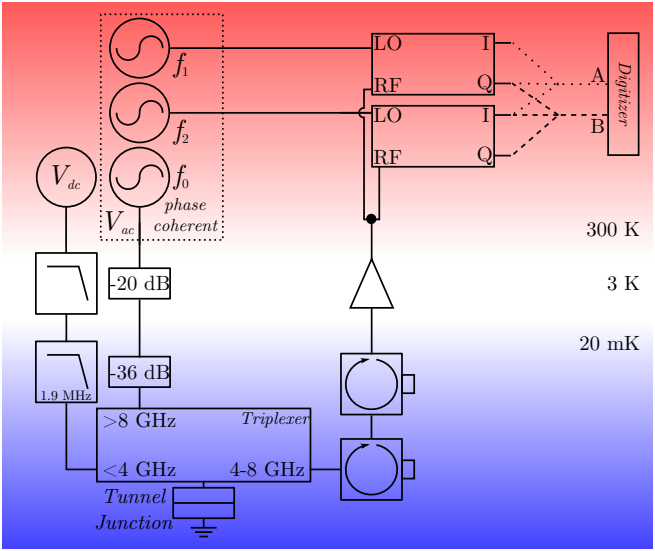}
    \caption{\footnotesize (color online) Experimental setup. See details in text.    \label{figMontage}}
\end{figure}

\emph{Experimental setup.} (see Fig. \ref{figMontage}).
We use a $R=70\unit{\Omega}$ Al/Al$_2$O$_3$/Al tunnel junction in the presence of a magnetic field to insure that the aluminum remains a normal metal at all temperatures. It is cooled to $\sim18\unit{mK}$ by a dilution refrigerator. A triplexer connected to the junction separates the frequency spectrum in three bands corresponding to the dc bias ($<4\unit{GHz}$), the ac bias at frequency $f_0$ ($>8\unit{GHz}$) and the detection band ($4-8\unit{GHz}$). Low-pass filters are used to minimize the parasitic noise coming down the dc bias line and attenuators are placed on the ac bias line to dampen noise generated by room-temperature electronics. The signal generated by the junction in the $4-8\unit{GHz}$ range goes through two circulators, used to isolate the sample from the amplification line noise, and is then amplified by a high electron mobility transistor amplifier placed at $3\unit{K}$.

At room temperature, the signal is separated in two branches, each entering an IQ mixer. One mixer is referenced to a local oscillator at frequency $f_1$ and the other at frequency $f_2=f_0-f_1$. All three microwave sources at frequencies $f_0$, $f_1$ and $f_2$ are phase coherent. The two IQ mixers take the signal emitted by the junction and provide the in-phase $X_{1,2}$ and quadrature $P_{1,2}$ parts relative to their references with a bandwidth of $80\unit{MHz}$. Similar setups have already been used to determine statistical properties of radiation in the microwave domain \cite{Menzel2010, Mariantoni2010, Menzel2012,Bozyigit2011,Eichler2011}. We chose to work at $f_0=14.5\unit{GHz}$, $f_1=7\unit{GHz}\Rightarrow f_2=7.5\unit{GHz}$ so that $f_1$ and $f_2$ are far enough to suppress any overlap between the frequency bands. Any two quantities $A,B$ among $X_1$, $X_2$, $P_1$ and $P_2$ can be digitized simultaneously by a two-channel digitizer at a rate of $400\unit{MS/s}$, yielding a 2D probability map $\mathcal{P}\inp{A,B}$.   From $\mathcal{P}\inp{A,B}$, one can calculate any statistical quantity, in particular the variances $\moy{A^2}$, $\moy{B^2}$, as well as the correlators $\moy{AB}$.

\emph{Calibration.}
The four detection channels must be calibrated separately. This is achieved by measuring the variances $\moy{X_{1,2}^2}$, $\moy{P_{1,2}^2}$ in the absence of RF excitation. These should all be proportional to the noise spectral density of a tunnel junction af frequency $f_{1,2}$, given by $S\inp{f,V}=\inp{S_0\inp{f+eV/h}+S_0\inp{f-eV/h}}/2$ where $S_0\inp{f}=\inp{hf/R}\coth\inp{hf/2k_BT}$ is the equilibrium noise spectral density at frequency $f$ in a tunnel junction of resistance $R$. By fitting the measurements with this formula, we find an electron temperature of $T=18\unit{mK}$ and an amplifier noise temperature of $\sim 3\unit{K}$, which are identical for all four channels. The small channel cross-talk was eliminated using the fact that $\moy{A_1B_2}=0$ when no microwave excitation is present.

\begin{figure}
    \includegraphics[width=\columnwidth] {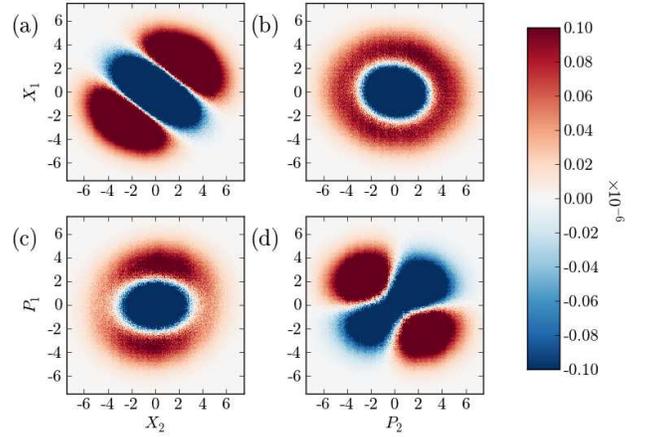}
    \caption{\footnotesize (color online) $\Delta \mathcal{P}= \mathcal{P}\inp{A,B}_{\vdc,\vac}- \mathcal{P}\inp{A,B}_{0,0}$ (unitless): (a)-(d) Difference between the normalised 2D current fluctuation distributions of the tunnel junction  at $\vdc=29.4\unit{\mu V}$, $18\unit{mK}$, under $14.5\unit{GHz}$, $\vac=37\unit{\mu V}$ microwave excitation and without excitation at $\vdc=0$. $X$ and $P$ represent 2 quadratures of the observed signal while numbers 1 and 2 represent fluctuations of frequency $7\unit{GHz}$ and $7.5\unit{GHz}$ in arbitrary units. Each distribution is made up of $2\ee{11}\unit{samples}$.
    \label{colMap}}
\end{figure}

\emph{Results.}
We begin our analysis by illustrating the effect of microwave excitation and dc bias on the current fluctuations.  We show on Fig. \ref{colMap} the difference $\Delta \mathcal{P}$ between the probability distribution $\mathcal{P}\inp{A,B}$ in the presence of  $\vdc=29.4\unit{\mu V}$ and $\vac=37\unit{\mu V}$ bias and $\mathcal{P}\inp{A,B}$ for $\vdc=\vac=0$. 
Fig. \ref{colMap}(b) and (c), which correspond to $\Delta \mathcal{P}\inp{X_1,P_2}$ and $\Delta \mathcal{P}\inp{P_1,X_2}$, are almost invariant by rotation. This means that the corresponding probability $\mathcal{P}\inp{X,P}$ depends only on $X^2+P^2$. 

As an immediate consequence, one expects $\moy{X_1P_2}=\moy{P_1X_2}=0$: $X_1$ and $P_2$ are uncorrelated, as are $X_2$ and $P_1$. In contrast, Fig. \ref{colMap}(a) and (d), which show respectively $\Delta \mathcal{P}\inp{X_1,X_2}$ and $\Delta \mathcal{P}\inp{P_1,P_2}$, are not invariant by rotation: the axes $X_1=\pm X_2$ and $P_1=\pm P_2$ are singular: for a given value of $X_1$ the probability $\Delta \mathcal{P}\inp{X_1,X_2}$ is either maximum or minimum for $X_2=X_1$, whereas for a given value of $P_1$, the probability $\Delta \mathcal{P}\inp{P_1,P_2}$ is either maximum or minimum for $P_2=-P_1$. These indicate that it is possible to observe correlations or anticorrelations  between $X_1$ and $X_2$ on one hand, and between $P_1$ and $P_2$ on the other hand. Data in Fig. \ref{colMap}(a) through (d) correspond to two frequencies $f_1$ and $f_2$ that sum up to $f_0$, all three frequencies being phase coherent. If this condition is not fulfilled, no correlations are observed between any two quadratures, giving plots similar to Fig. \ref{colMap}(b) or (c) (data not shown). The effect of frequencies on correlations between power fluctuations has been thoroughly studied in Refs. \cite{C4classique,C4quantique}.

\begin{figure}
    \includegraphics[width=\columnwidth] {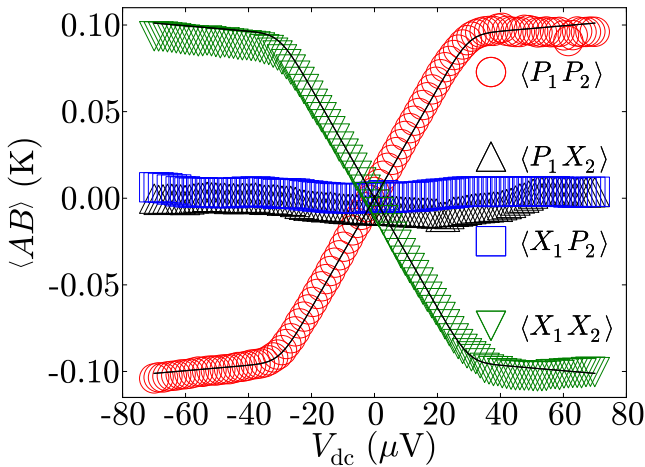}
    \caption{\footnotesize (color online) Quadrature correlators of EM field at frequencies $f_1=7\unit{GHz}$, $f_2=7.5\unit{GHz}$ generated by a $70\unit{\Omega}$ tunnel junction at $18\unit{mK}$ under $14.5\unit{GHz}$, $\vac=37\unit{\mu V}$ microwave excitation. Symbols represent experimental data and lines are theoretical expectations. Symbol sizes represent experimental uncertainty.
    \label{I1I2Q1Q2}}
\end{figure}

To be more quantitative, we show on Fig. \ref{I1I2Q1Q2} the $\moy{AB}$ correlators as a function of the dc bias voltage for a fixed $\vac$. Clearly, $\moy{X_1P_2}=\moy{P_1X_2}=0$ while $\moy{X_1X_2}=-\moy{P_1P_2}$ is non-zero for $\vdc\neq0$. These results are presented in temperature units (K), using the usual unit conversion $T_{noise}=RS/2k_B$ for the measured noise spectral density $S$ of a conductor of resistance $R$.

\emph{Theory.}
We now compare our experimental results with theoretical predictions. In order to link the measured quantities to electronic properties, we will first define the quadrature operators, following Ref. \cite{PAN_squeezing,Qassemi}
\begin{equation}
\hat{X}_{1,2}=\frac{\hat{I}\inp{f_{1,2}}+\hat{I}\inp{-f_{1,2}}}{\sqrt{2}},\;\;
\hat{P}_{1,2}=\frac{\hat{I}\inp{f_{1,2}}-\hat{I}\inp{-f_{1,2}}}{i\sqrt{2}}\nonumber
\label{eqXP}
\end{equation}
where $\hat{I}\inp{f}$ is the current operator at frequency $f$. In the absence of RF excitation, the currents observed at two different frequencies are uncorrelated, $ \moy{\hat{I}\inp{\pm f_1} \hat{I}\inp{\pm f_2}}=0$. The excitation at frequency $f_0=f_1+f_2$ induces correlations so that $\moy{\hat{I}\inp{f_1}\hat{I}\inp{f_2}}= \moy{\hat{I}\inp{-f_1}\hat{I}\inp{-f_2}}\neq0$.

More precisely, one has \cite{GR1,GR2,GR3}:

\begin{equation}
\moy{\hat{I}\inp{f}\hat{I}\inp{f_0-f}}=\sum_n \frac{\alpha_n}{2}[S_0\inp{f_{n+}}-S_0\inp{f_{n-}}]
\label{eqX}
\end{equation}
where $f_{n\pm}=f+nf_0\pm e\vdc/h$ and $\alpha_n=J_n\inp{e\vac/hf_0} J_{n+1}\inp{e\vac/hf_0}$, with $J_n$, the Bessel functions of the first kind. From this we can calculate the theoretical predictions for all the correlators, which are represented as black lines on Fig. \ref{I1I2Q1Q2}, showing a very good agreement between theory and experiment.

\begin{figure}
    \includegraphics[width=\columnwidth] {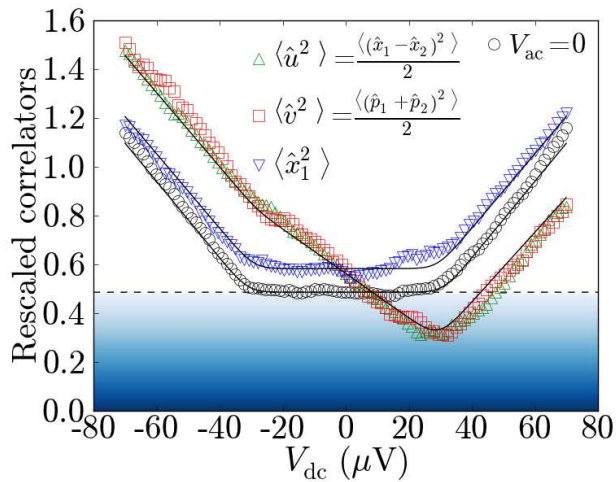}
    \caption{\footnotesize (color online)  Rescaled (unitless) variances of the EM field generated by a $70\unit{\Omega}$ tunnel junction at $18\unit{mK}$ under $14.5\unit{GHz}$, $\vac=37\unit{\mu V}$ microwave excitation and without excitation, obtained using signal quadratures at frequencies $f_1=7\unit{GHz}$, $f_2=7.5\unit{GHz}$. Symbols represent experimental data and lines are theoretical expectations. Symbol sizes represent experimental uncertainty. The shaded area showcases the less-than-vacuum noise levels.
    \label{mixData}}
\end{figure}

\emph{Two-mode squeezing.}
Squeezing is best explained using the dimensionless operators

\begin{equation}
\hat{x}_k =\frac{\hat{X}_k}{\sqrt{2hf_k/R}},\;\;\hat{p}_k=\frac{\hat{P}_k}{\sqrt{2hf_k/R}},
\nonumber
\label{eqPxp}
\end{equation}
chosen so that $\moy{\insb{\hat x_k, \hat p_{k'} }}=\delta_{k,k'}$ with $k, k'=1,2$. At equilibrium $\inp{\vdc=\vac=0}$ and very low temperature $\inp{k_BT\ll hf_{1,2}}$, the noise generated by the junction corresponds to vacuum fluctuations and $\moy{\hat{x}_k^2}=\moy{\hat{p}_k^2}=1/2$. Experimentally, this can be seen on Fig. \ref{mixData} as a plateau in $\moy{\hat{x}_k^2}$ vs $\vdc$ at $\vac=0$ (black circles) and low $\vdc$, which is outlined by a dashed line.

Single-mode squeezing refers to the possibility of going below $1/2$ for either $\moy{\hat{x}_k^2}$ or $\moy{\hat{p}_k^2}$. This has been observed when the excitation frequency $f_0$ corresponds to $f_0=f_k$ or $f_0=2f_k$\cite{PAN_squeezing}. As can be seen in Fig. \ref{mixData}, there is no single-mode squeezing for $f_0=f_1+f_2$ and $\moy{\hat{x}_1^2}>1/2$ (as well as $\moy{\hat{x}_2^2}=\moy{\hat{p}_1^2}=\moy{\hat{p}_2^2}>1/2$, data not shown). Using the operators $\hat{u}=\inp{\hat{x}_1-\hat{x}_2}/\sqrt{2}$ and $\hat{v}=\inp{\hat{p}_1+\hat{p}_2}/\sqrt{2}$, two-mode squeezing refers to the possibility of either $\moy{\hat{u}^2}$ or $\moy{\hat{v}^2}$ going below $1/2$, which corresponds to their value for vacuum fluctuations. As evidenced by Fig. \ref{mixData}, $\moy{\hat{u}^2}\simeq\moy{\hat{v}^2}$ goes below $1/2$ for a certain range of dc bias, a clear proof of the existence of two-mode squeezing for electronic shot noise.

The optimal squeezing observed corresponds to $\moy{\hat{u}^2}=0.32\pm0.05$, $\moy{\hat{v}^2}=0.31\pm0.05$, i.e. $2.1\unit{dB}$ below vacuum, versus the theoretical expectation of $\moy{\hat{u}^2}=\moy{\hat{v}^2}=0.33$. This minimum is observed at  $\vdc\simeq30\unit{\mu V}\simeq hf_{1,2}/e$. All data are in good agreement with theoretical predictions, plotted as full black lines on Fig. \ref{mixData} with $\moy{\hat{x}_k^2}=\moy{\hat{p}_k^2}=S\inp{f_k}$, the photo-assisted shot noise given by $S\inp{f}=\frac{1}{2}\Sigma_n J_n^2(e\vac/hf_0)\insb{S_0\inp{f_{n+}}+S_0\inp{f_{n-}}}$\cite{Lesovik}. Curves for $\moy{\inp{\hat x_1+\hat x _2}^2}/2$ and $\moy{\inp{\hat p_1-\hat p _2}^2}/2$ follow the same behaviour with reversed dc bias, showing minima of $0.35\pm0.05$ and $0.41\pm0.05$ or $1.6\unit{dB}$ at $\vdc\simeq -30\unit{\mu V}\simeq -hf_{1,2}/e$. The latter data was omitted from Fig. \ref{mixData} in order to simplify it.

\emph{Entanglement.}
While the presence of two-mode squeezing shows the existence of strong correlations between quadratures of the electromagnetic field at different frequencies, this is not enough to prove the existence of entanglement. A criterion certifying the inseparability of the two modes, and thus entanglement between them, is given in terms of the quantity $\delta=\moy{\hat{u}^2}+\moy{\hat{v}^2}$. In the case of a classical field, this must obey $\delta>1$\cite{Duan2000}. This is equivalent to a Bell-like inequality for continuous variables. As we reported in Fig. \ref{mixData}, we observe $\delta=0.6\pm0.1$. Thus, photons emitted at frequencies $f_1$ and $f_2$ are not only correlated but also form EPR pairs suitable for quantum information processing with continuous variables\cite{Braunstein2005}.

Two-mode quadrature-squeezed states are usually characterized by their covariance matrix. Following the notations of Ref. \cite{Giedke2003}, our experiment corresponds to $n=2\moy{x_{1,2}^2}\simeq2\moy{p_{1,2}^2}$, $k=2\moy{x_1x_2}\simeq-2\moy{p_1p_2}$ so that $\delta=n-k$. Equilibrium at $T=0$ corresponds to $n=1$ and $k=0$. Our observed optimal squeezing corresponds to $n=1.3\pm0.1$ and $k=0.52\pm0.05$. From these numbers, one can calculate all the statistical properties that characterize the electromagnetic field generated by the junction.  In particular, we find an formation entanglement of $E_F\simeq0.2$ (as defined in Ref. \cite{Giedke2003}) and a purity of $\mu=0.82$ (as defined in Ref. \cite{DiGuglielmo2007}). While in our experiment, the entangled photons are not spatially separated, this could easily be achieved using a diplexer, which can separate frequency bands without dissipation.

A cursory analysis of the data in terms of EPR steering shows that according to the definitions presented in Ref. \cite{Cavalcanti2009}, we obtain $\mu'=0.62$ (not to be confused with $\mu$ from Ref. \cite{DiGuglielmo2007}), $\overline{n}=0.15$. According to Eq. (71) and (77) from that work, we clearly respect the criteria for entanglement but fall shy of being able to observe steering. Numerical calculations using Eq. \ref{eqX} show that the condition for steerability could be fulfilled at lower temperature ($T<16.8\unit{mK}$ with the present setup) or higher frequency, while we should observe $\mu'=0.85$ and $\overline{n}=0.06$ at $T=0\unit{K}$. Although a temperature of $T<16.8\unit{mK}$ seems experimentally accessible, the physical meaning of steerability in terms of electron quantum shot noise still needs to be understood.

We acknowledge fruitful discussions with A. Bednorz, W. Belzig, A. Blais, G. Gasse and F. Qassemi. We thank L. Spietz for providing us with the tunnel junction and G. Lalibert\'e for technical help. We are also grateful to J. Schneeloch for pointing us towards the phenomenon of steerability. This work was supported by the Canada Excellence Research Chairs program, the NSERC, the MDEIE, the FRQNT via the INTRIQ, the Universit\'{e} de Sherbrooke via the EPIQ, and the Canada Foundation for Innovation.

\end{document}